\documentstyle[11pt,fullpage,doublespace,epsf]{article}
 \DeclareMathSizes{11}{19}{13}{9}   

\makeatletter
\@addtoreset{equation}{section}
\makeatother

\begin{document}

  \title{Graphics Turing Test}
\author{Michael McGuigan\\Brookhaven National Laboratory\\Upton NY 11973\\mcguigan@bnl.gov}
\date{}
\maketitle

\begin{abstract}We define a Graphics Turing Test to measure graphics performance in a similar manner to the definition of the traditional Turing Test. To pass the test one needs to reach a computational scale, the Graphics Turing Scale, for which Computer Generated Imagery becomes comparatively indistinguishable from real images while also being interactive. We derive an estimate for this computational scale which, although large, is within reach of todays supercomputers. We consider advantages and disadvantages of various computer systems designed to pass the Graphics Turing Test. Finally we discuss commercial applications from the creation of such a system, in particular Interactive Cinema.
\end{abstract}

Defining graphics computing scales can be difficult. Traditional methods such as counting triangles drawn per second can be misleading especially if a large percentage of the computing power is going into shading code or a physics simulation which can be involved in  making the  graphics image realistic. 

More than fifty years ago Alan Turing faced a similar challenge in evaluating the power of a computer. In short, he devised a test to see if one can build a computer whose intelligence was indistinguishable from human intelligence \cite{Turing}. The Turing computational scale is then defined by the computational power required to pass this test. The beauty of this definition is that the computational scale is defined by the indistinguishability of comparison. In particular one does not need a definition of human intelligence to implement the test. Also one does not need to measure in terms of Flops per second or some other commonly used method of measuring computer power, after all integer performance, memory access or network connectivity may be important. One only needs to statistically compare how subjects can discriminate between a human and a computer. If the subjects can do no better than a random guess than the Turing test is passed and the Turing scale has been measured. Unfortunately the Turing Test is too hard for now, one cannot even use partial successes in AI to estimate the size of a computer required to pass the test \cite{Turing2}. As we shall see the corresponding situation in graphics is somewhat better.

In a similar approach to the traditional Turing test we define the Graphics Turing Test as follows:
\\
\\
{\it The subject views and interacts with a real or computer generated scene. The test is passed if the subject can not determine reality from simulated reality better than a random guess.
(a) The subject operates a remotely controlled ( or simulated) robotic arm and views a computer screen.
(b) The subject enters a door to a controlled vehicle or motion simulator with computer screens for windows. An eye patch can be worn on one eye as stereo vision  is difficult to simulate.}
\\
\\
The Graphics Turing Scale is then defined as the computer power necessary to pass the above test. The key feature in the above definition is that the subject interacts with the computer generated scene. As we shall discuss it is possible using a reasonably powerful system to create a computer generated image that is indistinguishable from reality, but it may take several hours to render the image. It is the requirement of interactivity that accounts for the large amount of computer power inherent in passing the Graphics Turing Test. Note that, as with the traditional Turing Test, in the Graphics Turing Test the computational scale is defined as one reaches an indistinguishability of comparison. Also note that one need not define the graphics performance in terms of triangles per second or pixel fill rate or some other commonly used metric. The computational scale is defined intrinsically by the subjects ability to determine if the scene is real or computer generated even by interacting and driving through the scene. It is the statistical analysis of the subject's determination of the reality of the scene compared to a random guess that forms the metric in  the Graphics Turing Test. All traditional measures of graphics power including the complexity of the geometry, shader code and lighting has already been folded into this metric. Some specific implementations of the Graphics Turing Test which are relatively easy to set up are stated in (a) and (b) above. Other realizations are possible. In particular, if ghosting can be sufficiently eliminated, it should be possible to setup a stereoscopic version of the Graphics Turing Test. It has also been proposed to combine realistic graphics into the traditional Turing Test \cite{Ultimate}. However, as it is discussed in this paper, the Graphic Turing Test measures graphics performance only and is separate from AI.

But what is this Graphics Turing Scale and what type of computer is required to pass the Graphics Turing Test? Although the Graphics Turing Test is similar in structure to the traditional Turing Test it differs in one important aspect. Partial success in generating non interactive photo realistic imagery can be used to estimate the Graphics Turing Scale. The large computing scale is mainly the result of achieving an interactive frame rate for the computer generated imagery, so that the subject perceives time as continuous, and in implementing relatively mature graphics algorithms. Although a hard problem the Graphics Turing Test is within reach of todays supercomputers. In contrast, for the traditional Turing Test the AI algorithms are less mature and non interactive AI has not been achieved.

To estimate the size of the Graphics Turing Scale consider the recent photo realistic renderings by Paul Debevec of the Parthenon \cite{Parthenon}. He used Monte Carlo illumination methods and about 1GB of complex geometry. The results were essentially indistinguishable from reality which is all the more impressive as they were presented to an audience of graphics professionals acutely aware of the subtleties of implementing realistic lighting. To generate the imagery required 2 hours on 1 CPU 2.4 Ghz Pentium IV for each frame. Thus an interactivity of 1/30 sec would require 216,000 CPUs of computing power. Although more powerful CPUs are available today it is also true that supercomputers typically use lower clock rates to reduce heat and power consumption as thousands of processors are placed in close proximity to one another. Thus we estimate the Graphics Turing Scale as the computational equivalent of roughly 200k CPUs. In terms of traditional measures of computing power we have:
\[
{\rm Graphics\; Turing\; Scale } = (216\;\rm{k}) \times (4.8{\rm  \;GFlops)} \times \left\{ {\begin{array}{*{20}c}
   1  \\
   \epsilon   \\
\end{array}} \right\} = \left\{ {\begin{array}{*{20}c}
   {1036.8{\rm  \;TFlops\; Peak}}  \\
   {518.4{\rm  \;TFlops\; Sustained}}  \\
\end{array}} \right\}
\]
where $\rm{GFlops}$ and $\rm{TFlops}$ are billion and trillion floating point operations per second respectively and $\epsilon$ is the computational efficiency of the rendering algorithm which we take to be 50\%. Peak refers to the theoretical computing power attainable in a system and Sustained refers to the computer power attained when taking into account the inefficiency of the algorithm. The efficiency is given by the ratio $\epsilon = {\rm Sustained/Peak}$.

One way to  achieve a 200k times speedup in graphics rendering is through a interactive parallel graphics implementation. Another is through a large render farm although this would most likely be non interactive.
As indicated above most interactive parallel graphics implementations are not 100\% efficient \cite{Tomov1}. This because the communication among the processors in distributing the geometry or in assembling the final image causes inefficiency when the processors are waiting for data to be sent or received, before they can compute. However using efficient message passing protocols and possibly using assembler programming 50\% efficiency can be achieved. An interactive parallel rendering system with 400k processors and low latency to minimize the inefficiency in processor communication would seem to be the type of system most likely to pass the Graphics Turing Test.

\pagebreak
We consider a variety of systems and their applicability to the Graphics Turing Test:

(1) {\bf Graphics Grid}. If 1 million people connect their computers together in a Graphics Grid one has more than enough rendering power to create 2 hours of photo realistic imagery in 2 hours wall clock time with the Grid operating as an extremely large render farm. The difficulty here is in achieving interactive frame rates due to the slow communication time between computers. This is because the interprocessor communication on a Grid system is determined by the network and these times are longer than those found on today supercomputers which are highly localized. Nevertheless some interactivity can be input into the rendering process and it is difficult to beat the low cost of this system as each participant provides and maintains their own computer.

(2) {\bf Graphics Cluster}. A graphic cluster resembles an ordinary compute cluster with the addition of graphics cards containing GPUs that can be used for rendering as well as for physics based simulation. Current GPU based photo realistic rendering such as Gelato typically run twice as fast as CPU based renderers \cite{Gelato}. Physics based simulation that adds realism to CGI can also be run on the GPU with speedups of 3-5 times those found in a CPU based system \cite{Tomov2}\cite{Kaufman}. Fast connection between processors using an Infiniband interconnect would allow interactive parallel rendering as well. The difficulty in a system of this type is that only relatively small Graphics Cluster systems have been used so far, at least relative to the Graphics Turing Scale. The largest such system is about 256 GPUs. However the interactivity of the GPU Cluster system as well as output to large tiled display walls would make an ideal connection system between a subject and a supercomputer.

(3) {\bf Supercomputer}. Todays supercomputers contain thousands of processors and extreme low latency networks. For example the QCDOC system at BNL contains 12,288 1 GFlops processors and a six dimensional communication network with .2 microsecond latency \cite{Christ}.  
The Altix system at NASA contains 10,160 Itanium processors in a  combination shared memory infiniband interconnect. The IBM BlueGeneL system at Livermore contains 131,072 processors each with 2.8GFlops, .03 microsecond latency between nearest neighbors and .144 microsecond latency in all-to-all communication \cite{Gara}.  Only the last system IBM BlueGeneL with 367 TFlops peak ( 280.6 TFlops sustained) would seem to have enough compute power to pass the Graphics Turing Test. The last in the BlueGene series BlueGeneQ is expected to have 3000 TFlops peak would have more than enough to pass the Graphics Turing Test even stereoscopically. One drawback to supercomputer systems besides their large expense is that they are typically not run interactively. One submits a series of requests through a batch queuing system and sometime later receives the results. Nevertheless high data output can be achieved, usually through a large shared memory server which serves as an intermediary to move data off the supercomputer, and large scale parallel visualization can be considered on these systems \cite{Ahrens}.

(4) {\bf Combination System}. Perhaps the strongest approach to the building a system that can pass the Graphics Turing Test is through a  combined system of all of the above where a large supercomputer performs fast parallel rendering and outputs the image data to a graphics cluster which drives the displays of the motion simulator in scenario (b). The role of the Graphics Grid would then be used for input data from multiple users into the simulation. This aspect of the system is important for commercial applications as we now discuss. 

{\bf Commercial Applications}. Why would someone build a system to pass the Graphics Turing Test? Probably not to achieve an esoteric milestone in computer science, especially given the large cost of acquiring a supercomputer system. First note that synonymous with passing the Graphics Turing Test is  the achievement of the following: Artificial reality, Virtual Reality, Augmented Reality, Cinematic Gaming, Interactive Cinema. All these descriptions are essentially equivalent in terms of the graphics power that is required but each suggests a different application area. For example Virtual Reality  suggests applications to physically accurate simulators, Cinematic Gaming suggests realistic multiuser computer based games. 

We find the most intriguing application area to be Interactive Cinema which we discuss in more detail. When one finishes watching a movie one often has the feeling that they wished it could turn out differently. With multiple user input each theatrical  experience will turn out different depending on the input from the audience. Even if  the audience input is nearly uniform it is known from the butterfly affect that a small random input into the plot line of movie can have a large effect on the final outcome. This is the main commercial advantage of Interactive Cinema. A theater goer will plug their laptop into a network and interactively effect the plot line. Real time photo realistic rendering of the underlying system allows the movie to be created on the fly. This is essentially the appeal of video games, that they can be played over and over with different experiences each time, as opposed a fixed experience of the current cinema. By blending the two, one seeks a commercial advantage by combining the two markets. This represents a commercial application of the responsive medium concept of Myron Kreuger \cite{Kreuger}. The cost of the graphics system with 200k times the rendering capability of todays computers can be estimated at 500 million dollars. However the system can be used multiple times a day perhaps for several years. The market for interactive cinema would seem to be at least 700 million dollars based on current box office statistics and with an increase take due to multiple viewings. Thus economics would indicate an advantage to the creation of such a graphics system in the near future.

\end{document}